# Tunable high-Chern-number Chern insulators in rhombohedral tetralayer graphene/hBN moiré superlattices

Chuanqi Zheng[1,†], Chushan Li[1,†], Ke Huang[2,†], Chenyu Zhang[1], Kenji Watanabe[3], Takashi Taniguchi[4], Hao Yang[1,5,6], Dandan Guan[1,5,6], Liang Liu[1,5,6], Shiyong Wang[1,5,6], Yaoyi Li[1,5,6], Hao Zheng[1,5,6], Canhua Liu[1,5,6], Jinfeng Jia[1,5,6,7,8], Xueyang Song[9], Zhiwen Shi[1], Guorui Chen[1], Xiao Li[2*], Tingxin Li[1,5*] and Xiaoxue Liu[1,5,6*]

[1]State Key Laboratory of Micro-nano Engineering Science, Tsung-Dao Lee lnstitute & School of Physics and Astronomy, Key Laboratory of Artificial Structures and Quantum Control (Ministry of Education), Shanghai Jiao Tong University, Shanghai 200240, China

[2]Department of Physics, City University of Hong Kong, Kowloon, Hong Kong SAR, China

[3]Research Center for Electronic and Optical Materials, National Institute for Materials Science, 1-1 Namiki, Tsukuba, Japan

[4]Research Center for Materials Nanoarchitectonics, National Institute for Materials Science, 1-1 Namiki, Tsukuba, Japan

[5]Hefei National Laboratory, Hefei 230088, China

[6]Shanghai Research Center for Quantum Sciences, 99 Xiupu Road, Shanghai 201315, China3

[7]Quantum Science Center of Guangdong-Hong Kong-Macao Greater Bay Area (Guangdong), Shenzhen 518045, China

[8]Department of Physics, Southern University of Science and Technology, Shenzhen 518055, China

[9]Department of Physics, Hong Kong University of Science and Technology; Clear Water Bay, Hong Kong SAR, China

[†]These authors contributed equally to this work.

[*]Emails: xiao.li@cityu.edu.hk, txli89@sjtu.edu.cn, xxliu90@sjtu.edu.cn

## Abstract

Moiré superlattices based on rhombohedral multilayer graphene have emerged as a highly tunable platform for engineering correlated topological phases. Here, we systematically investigate the transport properties of the hole-doped side in rhombohedral tetralayer graphene/ hexagonal boron nitride (hBN) moiré superlattices across a range of twist angles and alignment orientations. Notably, we observed multiple high-Chern-number Chern insulators, including the previously reported integer Chern insulator with Chern number $C$ = -4 at moiré filling factor $\nu$ = -1 and newly discovered symmetry-broken Chern insulating states with $C$ = +3, ±2, ±1 at fractional moiré fillings of $\nu$ = −2.5 or -2.6. These Chern insulating states emerge in both hBN alignment, but exhibit a sensitive moiré wavelength dependence. Our findings demonstrate the exceptional tunability of these high-Chern-number states via moiré wavelength, displacement electric field and external magnetic field, underscoring the distinct topological landscape realized in hole-doped RTG/hBN moiré superlattices.

**Introduction**

Moiré flat bands with nonzero Chern numbers provide a fertile ground for exploring strongly correlated quantum phases. Experimentally, at integer moiré fillings (corresponding to an integer number of electrons or holes per moiré unit cell), orbital ferromagnetism and integer Chern insulators (ICIs) have been observed in a variety of moiré systems, including twisted graphene systems [1-7], semiconductor moiré heterobilayers and homobilayers [8-14], as well as rhombohedral multilayer graphene (RMG) aligned with hexagonal boron nitride (hBN) moiré systems [15-27]. At fractional moiré fillings, sufficiently strong electronic correlations can open charge gaps, giving rise to two distinct types of topological states: i) fractional Chern insulators (FCIs) [28-32], which exhibit fractionally quantized Hall conductance and fractionally charged excitations, representing lattice analogues of the fractional quantum Hall effect in the absence of Landau levels; and (ii) symmetry-broken Chern insulators (SBCIs) with integer-quantized Hall conductance, which can be viewed as topological charge-density-wave states with an enlarged real-space unit cell. The competition between these two scenarios depends sensitively on the underlying band quantum geometry. While most experimental efforts at fractional filled moiré Chern bands to date have focused on bands with Chern number $C = 1$ [5-7, 9-12, 16-24], higher-Chern-number flat bands are predicted [33-39] to host an even richer landscape of quantum phases, including multicomponent fractional states, exotic anyons, and unconventional symmetry-broken orders arising from the internal "Chern flavor" degree of freedom.

Recently, RMG-based moiré systems are found to support high-Chern-number moiré bands [40-45]. In this work, we systematically investigate the transport properties of the hole-doped rhombohedral tetralayer graphene (RTG)/hBN moiré superlattices with different alignment orientations and moiré wavelengths. Remarkably, multiple Chern insulating states with high Chern numbers are observed at both integer and fractional moiré filling factors $v$. In addition to the previously reported [20] ICI with $C = -4$ at $v = -1$, notably, several additional SBCIs with $C = \pm 1$, $\pm 2$, and $+3$ are found to emerge at fractional moiré fillings around $v = -2.5$ and $-2.6$. All these Chern insulating states emerge in the moiré-distant region on the hole-doped side of RTG/hBN within a relatively low displacement electric field range (-0.3 V/nm < $D$ < 0). This is in contrast to the electron-doped side, where the topological flat bands with $C = 1$ appear only under a strong $D$-field of about 1 V/nm [16-24]. We find that the formation of these hole-doped Chern insulating states are negligibly affected by the hBN alignment orientation, while they are sensitive to the moiré wavelengths: the $C = -4$ Chern insulator at $v = -1$ emerge only when the moiré wavelength is above 13 nm, while the newly observed SBCI states display a more intricate moiré wavelength dependence. Moreover, unlike the $C = -4$ ICI state at $v = -1$, the SBCI states are suppressed above applied perpendicular magnetic field of $B_\perp \approx$ 0.3-0.5 T, and their evolution with $B_\perp$ appears to be inconsistent with the Streda formula.

**Transport phase diagram of RTG/hBN moiré superlattice on the hole-doped side**

Figure 1(a) shows the schematic of the RTG/hBN moiré devices. The characteristics of the RTG/hBN devices for measurements are summarized in Table I, and the detailed fabrication procedures are described in a separate paper [22] and in Supplemental Materials. Figure 1(b) and 1(c) show the longitudinal resistivity $\rho_{xx}$ as a function of the moiré filling factor $v$ and the displacement electric field $D$, measured in two RTG/hBN devices with opposite hBN alignment orientations, but similar moiré wavelength $\lambda$ of 14.41 nm (Device 1) and 14.13 nm (Device 2), respectively. The alignment orientation refers to the relative positioning between the graphene and hBN layers at 0° or 180°. We follow the same labeling criterion established in previous studies [22-24], where the two orientations are denoted as $\xi = 0$ or 1. Earlier works [22,23] have shown that on the electron-doped side—particularly in the moiré-proximal regime—the transport properties are strongly affected by the hBN alignment orientation.

On the hole-doped side, however, we found that devices with opposite alignment orientations exhibit qualitatively similar transport features, only with subtle distinctions in the moiré-proximal region. Unless otherwise specified, the moiré-proximal region (moiré-distant region) for hole doping corresponds to the $D > 0$ ($D < 0$) region in all plots. As shown in Fig. 1(b) and 1(c), both devices show a satellite resistance peak at $v = -4$ within a low electric field range (with -0.07 V/nm $< D <$ 0.21 V/nm for Device 1; -0.08 V/nm $< D <$ 0.12 V/nm for Device 2) and correlated insulators at $v = -2$ (with 0.07 V/nm $< D <$ 0.32 V/nm, -0.04 V/nm $< D <$ -0.22 V/nm for Device 1; 0.05 V/nm $< D <$ 0.21 V/nm, -0.04 V/nm $< D <$ -0.22V/nm for Device 2). However, we find that the resistivity value at $v = -4$ still weakly depends on the alignment orientation. In the moiré-proximal region, the resistivity at $v = -4$ reaches the order of megaohms for $\xi = 0$, whereas it is only on the order of several tens of kΩ for $\xi = 1$ (also see Fig. S1). Figures 1(d) and 1(e) present zoomed-in maps of $\rho_{xx}$ and $\rho_{xy}$ as functions of $v$ and $D$, measured under a small magnetic field of $B_\perp = 0.05$ T for Device 2 (see Fig. S1 for data from other devices). Substantially enhanced Hall resistivity $\rho_{xy}$, accompanied by $\rho_{xx}$ dips, emerges near both $v = -1$ and the fractional moiré filling $v \approx -2.5$, demonstrating the breaking of time-reversal-symmetry (TRS) and the formation of Chern insulating states at these fillings. Notably, we find that all of these TRS-breaking states appear in both hBN alignment types, and in each case, they emerge only in the moiré-distant region, being absent in the moiré-proximal region, showing an apparent asymmetry with respect to $D = 0$.

In Fig. 1(f), we theoretically calculate the single-particle density of states (DOS) of RTG/hBN as a function of $v$ and interlayer potential difference $\Delta$ (see calculation methods in Supplemental Materials). For both alignment orientations, there are single-particle band gaps at $v = -4$ and around $D = 0$. Nonetheless, the band gap for $\xi = 0$ is much larger than that for $\xi = 1$, consistent with experimental observations. The calculations further show that the first moiré valence band carries a valley Chern number $C_v = 4$ at negative $\Delta$, demonstrating the presence of a tunable high-Chern-number moiré flat band in RTG/hBN systems (Fig. S2). Furthermore, as shown in Fig. 1(f), the single-particle DOS exhibits hot spots around $v = -2$ in both the moiré-proximal and moiré-distant regions, and around $v = -1$ only in the moire-distant region. Such enhanced DOS potentially trigger symmetry-breaking instabilities, leading to the correlated insulators at $v = -2$ and ICI states at $v = -1$ observed in the experiments. By contrast, the DOS around $v = -3$ is relatively low, which may partially account for the absence of a corresponding insulating state. This argument is further verified by our Hartree-Fock calculations (Fig. S2), in which the insulating states appear at $v = -1, -2$ but not at $v = -3$.

**$C = -4$ QAH effect at $v = -1$ and its moiré wavelength dependence**

We further examine moiré wavelength dependence of the transport phase diagrams. Figures 2(a), 2(b), and 2(e)-2(l) present $\rho_{xx}$ and $\rho_{xy}$ as functions of $v$ and $D$ under a small magnetic field for RTG/hBN moiré superlattices with various hBN alignment orientations and moiré wavelengths. Figure 2(c) and 2(d) show the temperature dependent magnetic hysteresis loops of $\rho_{xx}$ and $\rho_{xy}$ for Device 1 at $v = -1$ and $D = -0.228$ V/nm. Quantum anomalous Hall (QAH) effects with quantized $\rho_{xy}$ at $h/4e^2$ and vanishing $\rho_{xx}$ are observed at low-temperatures and zero magnetic field, establishing an ICI state with $C = -4$ at $v = -1$, with the $\rho_{xy}$ quantization persisting up to $T \approx 1.5$ K, consistent with previous studies [20]. The Chern number obtained from Středa formula fitting is also consistent with $C = -4$ (Fig. S3). The negative Chern number on the hole-doping side follows from the valley-polarized nature of the Chern insulator. At zero field, holes can polarize into either valley with equal energy; a finite field lifts this degeneracy. By the Středa formula, the number of states in a band shifts by $C_v \phi/\phi_0$. For the highest moiré valence mini band in the K valley ($C_v = 4$), capacity increases with field; while in K', it decreases. At $v = -1 - 4\phi/\phi_0$, K-polarized holes fit entirely within the enlarged highest moiré valence mini band, whereas K'-polarization would force occupation of remote bands at substantial energy cost — yielding a robust Chern insulator with $C = -4$. At $v = -1 + 4\phi/\phi_0$, however, either valley can accommodate the holes, so the insulating K'-polarized state competes with a metallic K-polarized one and no robust Chern insulator forms [46].

Notably, the $C = -4$ integer Chern insulating state emerges in both hBN alignment configurations but is sensitive to the moiré wavelength. As illustrated in Fig. S3, QAH effects with $\rho_{xy} = h/4e^2$ have been observed at $v = -1$ in Device 1-3, with $\lambda$ longer than 13.18 nm, and with both $\xi = 0$ and $\xi = 1$ alignment configurations. However, as the moiré wavelength decreases, the region in the $v$-$D$ phase diagram near $v = -1$ hosting the $C = -4$ Chern insulating state narrows (Fig. 2(a), 2(b), 2(e)-2(h)). In Device 4 and 5, where the $\lambda$ is shorter than 13 nm, the $C = -4$ integer Chern state is fully suppressed, and no anomalous Hall (AH) effect is observed near $v = -1$ (Figs. 2(i)-2(l)).

**Multiple SBCI states with high-Chern numbers**

We next focus on the TRS-breaking states emerging at fractional fillings between $v = -2$ and $-3$. In all five measured RTG/hBN devices, AH effects are observed within a narrow $D$-field range and approximately over the filling range $v \approx -2.4$ to $-2.8$. Similar to the $C = -4$ ICI state at $v = -1$, these fractional-filling states are also largely insensitive to the hBN alignment but show dependence on the moiré wavelenth; however, their evolution with $\lambda$ is more subtle. Notably, SBCIs, characterized by integer-quantized $\rho_{xy}$ and nearly vanishing $\rho_{xx}$, yet emanating from fractional moiré fillings, are observed in Device 1, 2, 3 and 5 (Fig. 2 and Fig. S4). The corresponding Chern numbers vary among different devices, as summarized in Table I. Among all measured devices, Device 4 ($\lambda = 13.00$ nm) exhibits a relatively weak AH response (Fig. S5), and no SBCI states are observed. In addition, compared with other devices, the correlated insulating states at $v = -2$ and the satellite resistance peak at $v = -4$ in Device 4 occupy a narrower $D$ range. Nevertheless, its electron-doped transport properties, as previously shown in a separate paper [22], are consistent with those of other devices in both moiré-proximal and moiré-distant regions. The FCI states are observed in the moiré-distant region on the electron-doped side at $v = 2/3$, $3/5$, and $4/7$ in Device 4, highlighting the high quality of the device. The underlying mechanism responsible for the absence of SBCI states on the hole-doping side in Device 4 remains to be further investigated.

Interestingly, the SBCI states observed in Device 1,2,3 and 5 exhibit a variety of integer Chern numbers, including $C = \pm 1, \pm 2$ and $C = 3$. Figures 3(a) and 3(b) show the the $\rho_{xx}$ and $\rho_{xy}$, respectively, as a function of $v$ and $B_\perp$ for Device 1 at $D = -0.105$ V/nm. Similar plots for Device 3 are shown in Figs. 3(c) and 3(d). In both devices, SBCIs with quantized $\rho_{xy} = h/3e^2$, accompanied by nearly vanishing $\rho_{xx}$ are observed emerging from $v = -2.6$. This $C = 3$ SBCI state persists down to nearly zero magnetic field, although a nearby ferromagnetic metal state with an AH signal of opposite sign appears to compete with the $C = 3$ SBCI state. Additionally, another SBCI with $C = 2$ emanating from $v = -2.5$ is also observed in Device 1, which begins to develop at $B_\perp > 0.2$ T. The $C = 2$ SBCI state is also observed in Device 2 (Fig. S6). We also notice that these SBCI states become fully suppressed when the applied $B_\perp$ exceeds approximetly 0.5 T, and the dispersion of the $\rho_{xx}$ dips and $\rho_{xy}$ plateaus in the $v$-$B_\perp$ phase space does not follow the Středa formula. Fitting their dispersions using the Středa formula of $n = CeB/h$ yields a much larger Chern number values (approximately 6 - 9), as compared to the quantized $\rho_{xy} = h/Ce^2$ with $C = 2, 3$.

Figures. 3(e) and 3(f) further show the temperature dependence of $\rho_{xx}$ and $\rho_{xy}$ as a function of $v$ for the $C = 3$ Chern insulator in Device 3, measured at a small magnetic field of $B_\perp = 0.01$ T and $D = -0.12$ V/nm. At low temperatures, a $\rho_{xy}$ plateau close to $h/3e^2$ emerges near $v = -2.6$, together with a $\rho_{xx}$ minimum. Both the local $\rho_{xx}$ dip and the $\rho_{xy}$ plateau are gradually suppressed with increasing temperature. Figures 3(g) and 3(h) further illustrate the magnetic hysteresis loops measured at $v = -2.6$ and $D = -0.15$ V/nm. The $\rho_{xy}$ is almost quantized at $h/3e^2$ with a residual $\rho_{xx}$

≈ 2 kohm at $B_\perp$ = 0 T and at the base temperature. The fitted activation gaps for SBCI states are smaller (~0.75K -1 K) compared to the ICI gap at $v$ = -1 (~ 2 K), as shown in Fig. S7.

Although the phenomena observed in Devices 1–3 are qualitatively similar, a richer set of SBCI states is observed in Device 5. Figures 4(a)-4(d) show the finely-scanned $\rho_{xx}$ and $\rho_{xy}$ maps as functions of $v$ and $D$ at $B_\perp$ = 0.075 T in Device 5. In particular, Figs. 4(a), 4(b) and Figs. 4(c), 4(d) correspond to scans performed with opposite filling-factor sweep directions. The AH signals observed within a narrow range of -0.10 V/nm < $D$ < -0.08 V/nm near $v$ = -2.6 exhibit opposite signs for upward and downward filling sweeps, demonstrating electric-field switching of the orbital magnetic moment. Similar behaviors are also observed in devices 1-3 in our measurements (See Fig. S4 and Fig. S8), consistent with previous reported results [20]. By combining results from both sweep directions, four SBCI states with $C$ = ±1, ±2 can be identified in Device 5. Figures 4(e)-4(f) and Fig. S9 show the $\rho_{xx}$ and $\rho_{xy}$ as a function of $v$ and $B_\perp$ at $D$ = -0.088 V/nm ($D$ = -0.118 V/nm). It can be seen that the $C$ = ±1 SBCI states originate from $v$ = -2.5, but occupy different regions of the $v$-$D$ phase space. In contrast, the $C$ = ±2 SBCI states originate from $v$ = -2.6, and are nearly degenerate energy in the $v$-$D$ phase space, with the favored ground state at zero magnetic field being highly sensitive to filling-factor sweep direction. These observations suggest that the $C$ = ±2 SBCI states are time-reversal counterparts of each other, whereas the $C$ = ±1 SBCI states are in two distinct phases unrelated by time-reversal symmetry. The $C$ = ±2 states gain a fine and $D$-dependent energy splitting at weak magnetic fields. A well-developed $C$ = 2 QAH effect is observed at $v$ = −2.6 and $B_\perp$ = 0, as shown in Figs. 4(g)–4(h). The SBCI states observed in Device 5 also do not follow the Středa formula.

**Discussions and Conclusions**

The emergence of ICI state at $v$ = -1 and multiple Chern insulating states with high Chern numbers near $v$ = -2.5 and -2.6 highlights a distinct regime in hole-doped RTG/hBN moiré superlattices. It is particularly noteworthy that these hole-doped high-Chern-number insulating states are absent in thicker rhombohedral multilayer graphene (e.g., see Fig. S10 for that in pentalayer graphene/hBN moiré superlattices), suggesting their unique sensitivity to the layer number. The RTG/hBN appears to provide an ideal platform where the interplay between band topology and correlation effects is optimized to stabilize these exotic states within a high-Chern-number moiré flat band.

An interesting observation is that the SBCI states deviate from the conventional Středa relation and are suppressed by relatively weak magnetic fields ($B$ < 0.5 T), raising open questions about the underlying mechanism. One possible explanation is that the SBCI states observed here may not be fully gapped in the bulk, but instead form a partially compressible electronic crystal state. Future measurements of the electronic compressibility would be crucial for clarifying this point. Another possibility is that even a small perpendicular magnetic field may induce a substantial reconstruction of the Fermi surface in this system, leading to the disappearance of the SBCI states above and their apparent deviation from the Středa relation.

Beyond their fundamental significance, these findings suggest a viable pathway toward developing functional topological devices. The realization of high-Chern-number QAH effects implies the existence of multiple dissipationless chiral edge channels at zero magnetic field, which are promising for developing low-power topological devices. In addition, the ability to switch between different Chern numbers ($C = \pm 1, \pm 2, +3$) via subtle adjustments in twist angle or displacement field underscores their exceptional tunability. Furthermore, the observation of electrical switching of magnetic moments in these high-Chern-number states provides a viable mechanism for non-volatile memory devices. Consequently, the hole-doped RTG/hBN system serves as a versatile playground for both exploring correlated topological phases and advancing the topological devices.

**Acknowledgement**

This work is supported by the National Key R&D Program of China (Nos. 2022YFA1402404, 2022YFA1402702, 2022YFA1405400, 2021YFA1401400, 2021YFA1400100, 2020YFA0309000, 2019YFA0308600, 2022YFA1402401, 2020YFA0309000), the National Natural Science Foundation of China (Nos. 12350403, 12374045, 12174249, 12174250, 12141404, 12350005, 12174248, 92265102), the Quantum Science and Technology-National Science and Technology Major Project (Nos. 2021ZD0302600, 2021ZD0302500), the Natural Science Foundation of Shanghai (No. 24LZ1401100, 24QA2703700), Shanghai Municipal Science and Technology Major Project (grant No.2019SHZDZX01) and the Shanghai Jiao Tong University 2030 Initiative Program.. C.L. is supported by T.D. Lee scholarship. X. Liu acknowledges "Shuguang Program" supported by Shanghai Education Development Foundation and Shanghai Municipal Education Commission. T.L. acknowledges support from the New Cornerstone Science Foundation through the XPLORER PRIZE. X. Li acknowledges the support from the Research Grants Council of Hong Kong (Grants No. 11312825, No. C7012-21G, and No. C7015-24G) and City University of Hong

Kong (Project No. 9610428). K. W. and T. T. acknowledge support from the JSPS KAKENHI (Nos. 21H05233 and 23H02052) and World Premier International Research Center Initiative (WPI), MEXT, Japan. A portion of this work was performed at the Synergetic Extreme Condition User Facility (SECUF).

# Main Figures

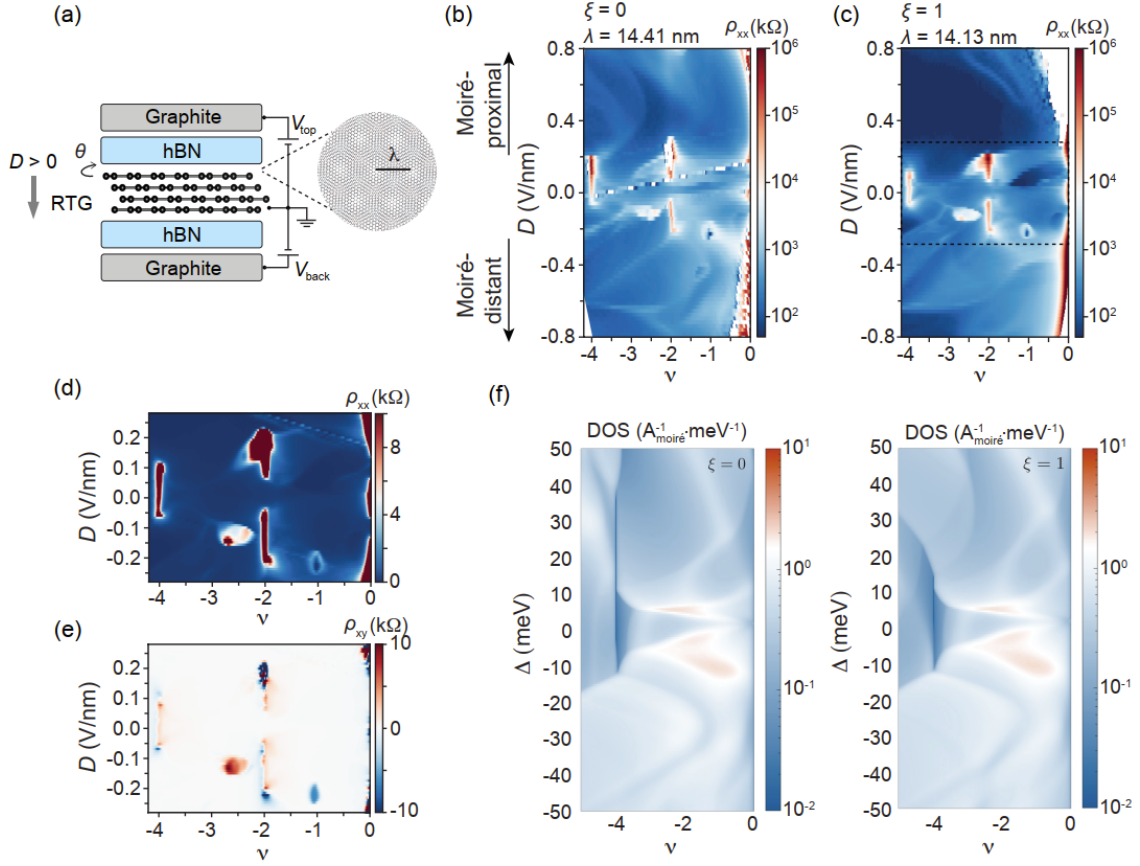

FIG. 1. Device characterization and transport phase diagrams of RTG/hBN moiré superlattices. (a) Schematic of the heterostructure, where the RTG layer is aligned with the top hBN layer to form a moiré superlattice with a moiré wavelength of $\lambda$. (b) and (c) Longitudinal resistivity $\rho_{xx}$ as a fucntion of $D$ and $\nu$, measured at $B_\perp = 0.1$T and $T = 15$ mK for devices 1(b) and 2 (c), respectively. (d) and (e) Phase diagrams of symmetrized $\rho_{xx}$ (d) and anti-symmetrized Hall resistivity $\rho_{xy}$ (e) measured at $B_\perp = \pm 0.05$T for the zoomed-in region of (c) indicated by dashed lines. (f) Single-particle DOS of RTG/hBN for two orientations. Here, we take a twist angle of 0.28° and consider a Gaussian broadening of 0.3 meV.

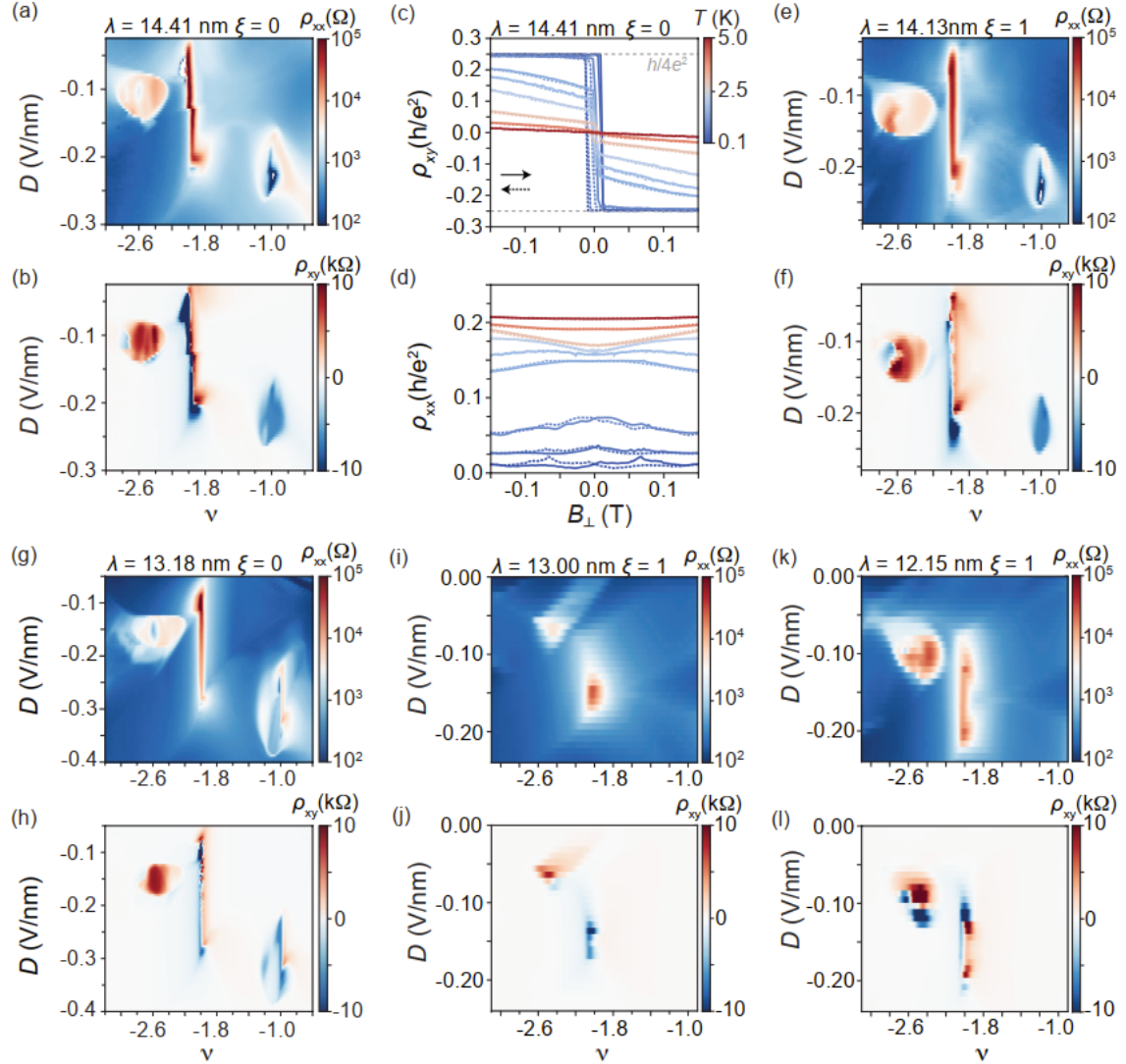

FIG. 2. TRS-breaking states in the moiré-distant region for hole doping. (a)-(b), (e)-(l) Phase diagrams of $\rho_{xx}$ and $\rho_{xy}$ measured at $B_\perp = 0.1$T and $T = 15$ mK for RTG/hBN devices with various twist angles and hBN alignment orientations. Device information is provided in the respective panels, and see Table I for detailed device characteristics. All data except for panels (a) and (b) are symmetrized (for $\rho_{xx}$) and anti-symmetrized (for $\rho_{xy}$) using measurements at $B_\perp = \pm 0.1$ T. (c) and (d) Temperature dependence of the magnetic hysteresis loops for $\rho_{xy}$ (c) and $\rho_{xx}$ (d), measured at $\nu = -1$ and $D = -0.228$ V/nm in device 1.

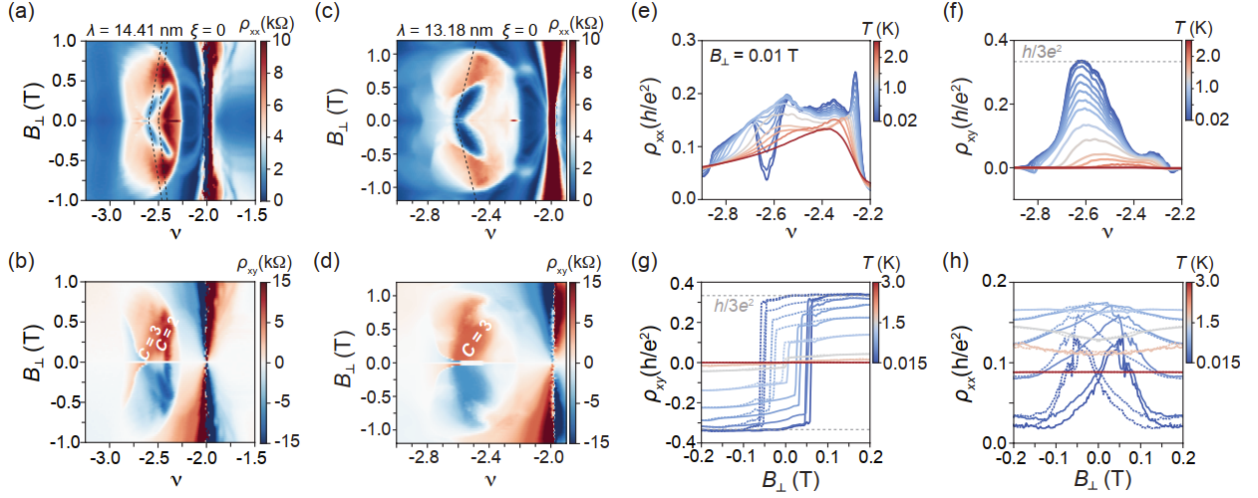

FIG. 3. Multiple Chern insulators with high Chern numbers at $\nu = -2.6$ and $-2.5$. (a) and (b) Landau fan diagram of $\rho_{xx}$ (a) and $\rho_{xy}$ (b) measured at $D = -0.105$ V/nm and $T = 15$ mK in device 1. The Chern insulators with $C = 2$ and $C = 3$ are indicated in (b). (c) and (d) Landau fan diagrams of $\rho_{xx}$ (c) and $\rho_{xy}$ (d) for device 3, measured at $D = -0.15$ V/nm and $T = 15$ mK. The Chern insulator with $C = 3$ is indicated in (d). As a guide to the eye, the black dashed lines in (a) and (c) represent slopes corresponding to Chern numbers $|C|$ according to the Streda formula. (e) and (f) Temperature dependence of symmetrized $\rho_{xx}$ and antisymmetrized $\rho_{xy}$ as a function of $\nu$, measured in device 3 at $B_\perp = \pm 0.01$ T and $D = -0.12$ V/nm. (g) and (h) Temperature dependence of the magnetic hysteresis loops for $\rho_{xy}$ (g) and $\rho_{xx}$ (h), measured in device 3 at $\nu = -2.6$ and $D = -0.15$ V/nm, for the $C = 3$ Chern state.

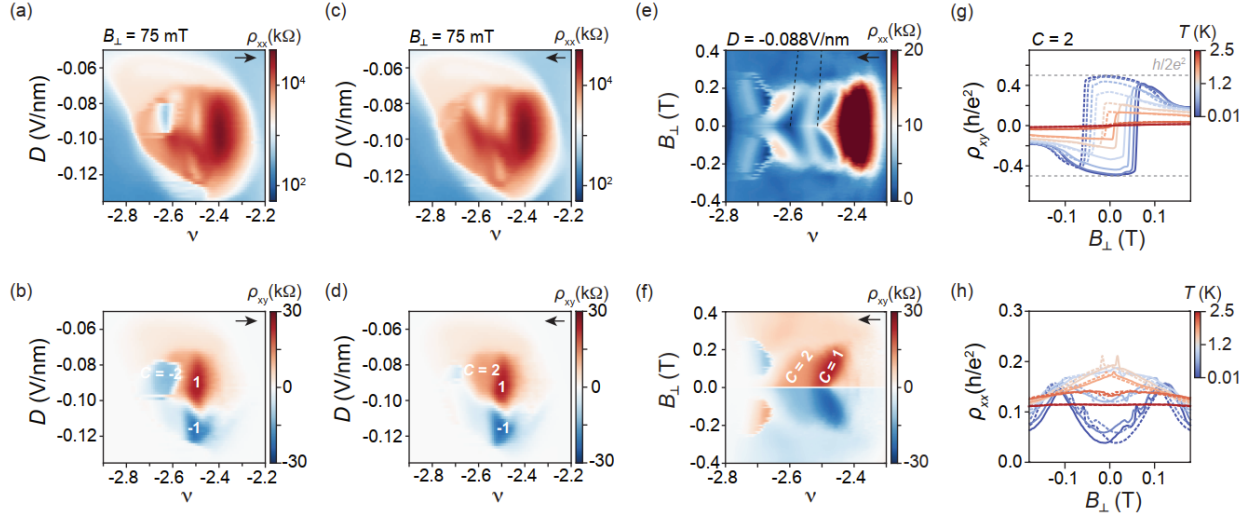

FIG. 4. Electric switching of magnetic moments for Chern states near $\nu = -2.6$. (a) and (b) Phase diagrams of symmetrized $\rho_{xx}$ (a) and antisymmetrized $\rho_{xy}$ (b) as a function of $\nu$ and $D$, measured at $B_\perp = \pm 75$ mT with sweeping the density in the upward direction. (c) and (d) Same measurements as in (a) - (b), but with sweeping the density in the downward direction. (e) and (f) Landau fan diagram of $\rho_{xx}$ (e) and $\rho_{xy}$ (f) measured at $D = -0.088$ V/nm with sweeping the density in the downward direction. As a guide to the eye, the black dashed lines in (e) represent slopes corresponding to Chern numbers $|C|$ according to the Streda formula. (g) and (h) Temperature dependence of the magnetic hysteresis loops for $\rho_{xy}$ (g) and $\rho_{xx}$ (h), measured at $\nu = -2.6$ and $D = -0.081$ V/nm, for the $C = 2$ Chern state. All data were measured on device 5 ($\lambda = 12.15$ nm). All panels except (g) and (h) were measured at $T = 15$ mK.

## Supplemental Materials

### Device Fabrication

Atomically thin flakes of graphite and hexagonal boron nitride (hBN) were obtained via mechanical exfoliation from bulk crystals. The rhombohedral stacking order of the graphene flakes was identified using near-field infrared microscopy and isolated from other stacking orders through AFM lithography (anodic oxidation). To improve device yield, the rhombohedral graphene was cut into multiple pieces using AFM lithography. All stacks were assembled using a standard dry-transfer technique with a poly(bisphenol A carbonate) (PC)/polydimethylsiloxane (PDMS) stamp. Specifically, a top hBN layer was picked up and used to sequentially pick up the rhombohedral graphene layers, which were then released onto a prepared bottom stack consisting of a bottom hBN layer and a bottom graphite gate on a Si/SiO$_2$ substrate. Near-field infrared microscopy was employed to identify regions where the rhombohedral graphene remained unrelaxed, and these areas were cleaned using AFM contact-mode before the top gate was released onto the corresponding position. Finally, the assembled stacks were patterned into a standard Hall bar geometry using e-beam lithography and reactive ion etching (CHF$_3$/O$_2$). Electrical edge contacts were finally fabricated by depositing Cr/Au (2/80 nm) films.

### Transport Measurement

Electrical transport measurements above 1.5 K were conducted in an Oxford Teslatron system. The measurements at temperatures below 1.5 K were performed in a cryogen-free dilution refrigerator (Q-one, Q-400) equipped with a 9 T superconducting magnet. Each fridge line has a sliver epoxy filter and multiple stage RC- filters at low temperature. The nominal base temperature is about 15 mK. A standard low-frequency lock-in technique with an excitation current of 1-3 nA at 3.77 Hz-23.37 Hz was used to measure longitudinal resistance $R_{xx}$ and Hall resistance $R_{xy}$.

We employed the following procedure to obtain the symmetrized $R_{xx}$ and antisymmetrized $R_{xy}$: $R_{xx}$ ($\pm B$) = [$R_{xx}$ (+B) + $R_{xx}$ (-B)]/2 and $R_{xy}$ ($\pm B$) = [$R_{xy}$ (+B) - $R_{xy}$ (-B)]/2. The measured $R_{xx}$ was then converted into longitudinal resistivity $\rho_{xx}$ using the formula of $\rho_{xx} = R_{xx}\frac{W}{L}$, where $W$ is the Hall bar width and $L$ is the separation between the voltage probes.

### Calibration of moiré filling factors

The dual-gate geometry of the device enables independent tuning of the carrier density $\left(n = \frac{c_t V_t + c_b V_b}{e} + n_0\right)$ and the perpendicular displacement electric field $\left(D = \frac{c_t V_t - c_b V_b}{2\varepsilon_0} + D_0\right)$ in the graphene layer by applying voltages to the top graphite gate ($V_t$) and bottom graphite gate ($V_b$). Here, $\varepsilon_0$, $c_t$, $c_b$, $n_0$ and $D_0$ represent the vacuum permittivity, the geometric capacitance of the top graphite gate, the geometric capacitance of the bottom graphite gate, the intrinsic doping and the built-in electric field, respectively. The geometric capacitance $c_t$ and $c_b$ were obtained from measuring quantum oscillations. The RTG/hBN moiré superlattice is formed by aligning the graphene with one hBN layer. The moiré wavelength $\lambda$ is determined from the Brown-Zak oscillation measurements (details provided in a separate paper [22]), and the corresponding twist angle $\theta$ is extracted using $\lambda = \frac{(1+\delta)a}{\sqrt{2(1+\delta)[1-\cos(\theta)]+\delta^2}}$. In this calculation, the lattice mismatch between hBN and graphene is taken as $\delta \approx 1.7\%$ and the graphene lattice constant as $a \approx 0.246$ nm. Detailed information for all devices obtained through this procedure is summarized in Table I (see also Ref. [22]).

**Theoretical Calculations**

**Single-particle theory for rhombohedral tetralayer graphene aligned with hBN**

The moiré system of rhombohedral multilayer graphene aligned with hBN has been extensively studied in Ref. [47-49]. In particular, recent experiments reveal that the orientation of the hBN plays a crucial role on the moiré proximal side, which can be theoretically attributed to the lattice relaxation and corrugation [49-51]. To incorporate these effects, we adopt the continuum model in Ref. [49, 52], given by

$$H^{(\xi)} = H_{RTG} + U^{(\xi)}(r),$$

where the effect of the hBN is described by $U^{(\xi)}(r)$, and $\xi = 0, 1$ defined in this work correspond to $\xi = -1, 1$ in Ref. [52], respectively. The Hamiltonian of RTG is composed of two parts, $H_{RTG} = h_0 + V_d$. The kinetic energy $h_0$ is given by,

$$h_0 = \begin{bmatrix} h^{(0)} & h^{(1)} & h^{(2)} & 0 \\ h^{(1)\dagger} & h^{(0)} & h^{(1)} & h^{(2)} \\ h^{(2)\dagger} & h^{(1)\dagger} & h^{(0)} & h^{(1)} \\ 0 & h^{(2)\dagger} & h^{(1)\dagger} & h^{(0)} \end{bmatrix},$$

where the Hamiltonian is written in the basis ($A_1$, $B_1$, $A_2$, $B_2$, $A_3$, $B_3$, $A_4$, $B_4$), and

$$h^{(0)} = -\frac{\sqrt{3}}{2}a_G t_0 \begin{bmatrix} 0 & k^\dagger \\ k & 0 \end{bmatrix}, \quad h^{(1)} = \begin{bmatrix} \frac{\sqrt{3}}{2}a_G t_4 k^\dagger & \frac{\sqrt{3}}{2}a_G t_3 k \\ t_1 & \frac{\sqrt{3}}{2}a_G t_4 k^\dagger \end{bmatrix}, \quad h^{(2)} = \begin{bmatrix} 0 & t_2/2 \\ 0 & 0 \end{bmatrix},$$

with $k = k_x + i k_y$. In addition, the potential of each layer and sublattice is

$$V_d = \Delta/2(-3, -3, -1, -1, 1, 1, 3, 3) + (0, \delta_1, \delta_2, \delta_2, \delta_2, \delta_2, \delta_1, 0),$$

where $\delta_1$ and $\delta_2$ are the onsite energy of corresponding sublattices, and $\Delta$ is the potential difference between two adjacent layers caused by the displacement field. Following [52], we take

$$(t_0, t_1, t_2, t_3, t_4, \delta_1, \delta_2) = (3100, 356.1, -8.3, 293, 144, 12.2, -16.4) \text{ meV}.$$

The moiré pattern of the system is generated by the lattice mismatch and twist angle between graphene and hBN. The reciprocal basis vectors of graphene are

$$G_1 = \frac{2\pi}{a_G}[1, -1/\sqrt{3}], \quad G_2 = \frac{2\pi}{a_G}[0, 2/\sqrt{3}],$$

and the reciprocal basis vectors of the aligned hBN are

$$G_i' = \frac{a_G}{a_{hBN}} R_\theta G_i$$

with a lattice mismatch $a_G/a_{hBN} \approx 1.018$. Here, $R_\phi$ is the counterclockwise rotation by angle $\phi$, and $\theta$ is the twist angle. Hence, the moiré reciprocal vectors are $g_i = G_i - G_i'$. The moiré potential is modeled by

$$U^{(\xi=0)}(r) = \begin{bmatrix} V_1 & V_2/\omega \\ V_2 & V_3 \end{bmatrix} e^{ig_1 \cdot r} + \begin{bmatrix} V_1 & V_2\omega \\ V_2\omega & V_3 \end{bmatrix} e^{ig_2 \cdot r} + \begin{bmatrix} V_1 & V_2 \\ V_2/\omega & V_3 \end{bmatrix} e^{-i(g_1+g_2) \cdot r} + \text{H.c.}$$

with $\omega = e^{2\pi i/3}$. To explain the experiment, we slightly adjust the parameters from [49, 52] and take

$$V_1 = -13.39\, e^{50.19°i} \text{ meV}, \quad V_2 = 11.34\, e^{19.60°i} \text{ meV}, \quad V_3 = 18.14\, e^{-46.64°i} \text{ meV}.$$

**Hartree-Fock theory**

For the Hartree-Fock calculation, we consider the following interaction,

$$V_{\text{int}} = \frac{1}{2A} \int dq\, V(q) {:} \rho(q)\rho(-q){:} - J \int dr {:} \vec{\sigma}_K(r) \cdot \vec{\sigma}_{K'}(r'){:}$$

where $A$ is the area of the system, the first term is the Coulomb potential, and the second term is the Hund's term [53]. Here, $\rho(q)$ is the Fourier transform of the total density of the four flavors, and $\vec{\sigma}_K(r)$, $\vec{\sigma}_{K'}(r')$ are the local spin operators of each valley. The gate-screened Coulomb potential $V(q)$ is given by

$$V(q) = \frac{2\pi e^2 \tanh qd}{\varepsilon q},$$

where $d = 30$ nm is the distance between the two gates, and the dielectric constant is taken to be $\varepsilon = 10$. Further, we take $J/A_{\text{moiré}} = 2$ meV, where $A_{\text{moiré}}$ is the area of the moiré unit cell. $J$ used here is comparable to the Hund's coupling used in Ref. [53]. The Hund's term corresponds to an energy decrease of $J/A_{\text{moiré}}$ per electron when there is one electron per moiré unit cell in both K ↑ and K′ ↓. Hence, $J > 0$ favors a spin-polarized state. Moreover, at odd-integer fillings, the occupied or unoccupied flavors also manifest an energy splitting of $J/A_{\text{moiré}}$. In Fig. S2, we show our Hartree-Fock calculations at $v$ = -1, -2, -3 and $\Delta$ = -12 meV. The interaction leads to flavor-polarized insulators at $v$ = -1, -2 for both orientations. In particular, the insulator at $v$ = -1 has a Chern number of $|C|$ = 4, in agreement with the experiments. However, the interaction is incapable of opening a global gap at $v$ = -3, resulting in a metallic state.

| Device | hBN alignment | Layer | Twist angle (degree) | Moiré wavelength (nm) | Chern insulators with $C = -4$ at $\nu = -1$ | Chern insulators near $\nu = -2.5$ (Chern number) |
|---|---|---|---|---|---|---|
| 1 | $\xi = 0$ | 4 | 0.20 | 14.41 | Yes | Yes ($C = 2, 3$) |
| 2 | $\xi = 1$ | 4 | 0.28 | 14.13 | Yes | Yes ($C = 2$) |
| 3 | $\xi = 0$ | 4 | 0.48 | 13.18 | Yes | Yes ($C = 3$) |
| 4 | $\xi = 1$ | 4 | 0.52 | 13.00 | No | No |
| 5 | $\xi = 1$ | 4 | 0.66 | 12.15 | No | Yes ($C = \pm 2, \pm 1$) |
| 6 | $\xi = 0$ | 5 | 0.20 | 14.41 | No | No |

**Table I. Summary of device characteristics**

Note: Some of these devices are reported in Ref. [22], but they have been renamed here for better clarity and consistency with the present study.

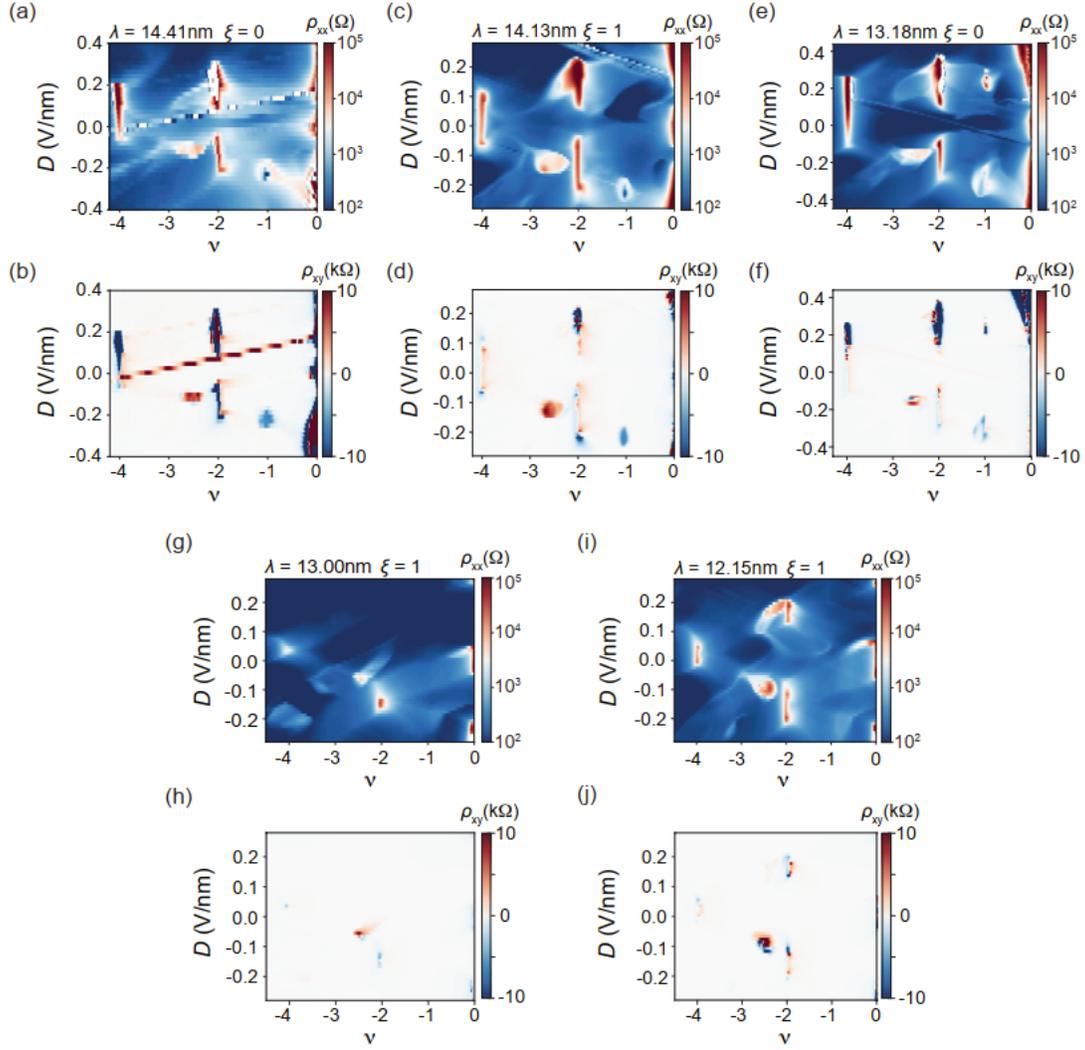

FIG. S1. Phase diagrams of RTG/hBN devices with various twist angles and hBN alignment orientations in the hole-doped regime. (a)-(j) Phase diagrams of $\rho_{xx}$ and $\rho_{xy}$ measured at $T = 15$ mK for RTG/hBN devices with various twist angles and hBN alignment orientations. Device information is provided in the respective panels. Except for panels (a) and (b), all data are symmetrized (for $\rho_{xx}$) and anti-symmetrized (for $\rho_{xy}$) using measurements at $B_\perp = \pm 0.05$ T. Panels (a) and (b) show measurements taken at $B_\perp = 0.1$ T.

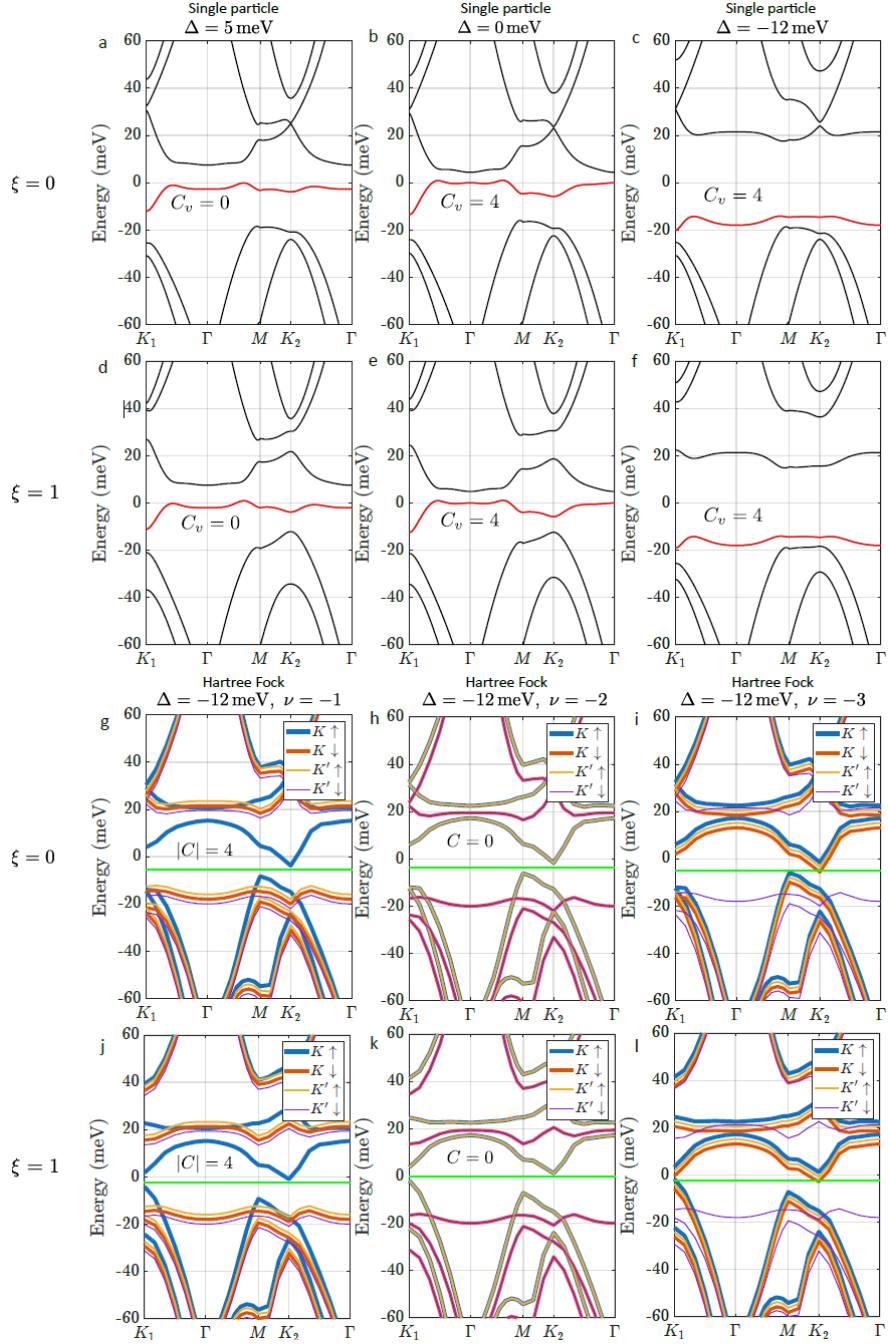

FIG. S2. Single-particle and Hartree-Fock band structure of RPG/hBN at $\nu$ = -1, -2, and -3. The bands in the $K$ valley are plotted along the path ($K_1$, $\Gamma$, $M$, $K_2$, $\Gamma$) and the bands in the $K'$ valley are plotted along the path ($-K_1$, $\Gamma$, $-M$, $-K_2$, $\Gamma$), so that the spectra of the four flavors exactly overlap in the absence of interactions. The first and third rows are for the $\xi = 0$ case, and the second and fourth for the $\xi = 1$ case. In (a-f), the red lines indicate the highest moiré valence mini band, with its valley Chern number given in each panel. In (g-l), the colors represent the four flavors, the horizontal green lines are the Fermi level, and $|C|$ denotes the Chern number of the ground state. The calculations in all panels are performed at $\theta = 0.28°$. Panels g-l are calculated at $\Delta = -12$ meV.

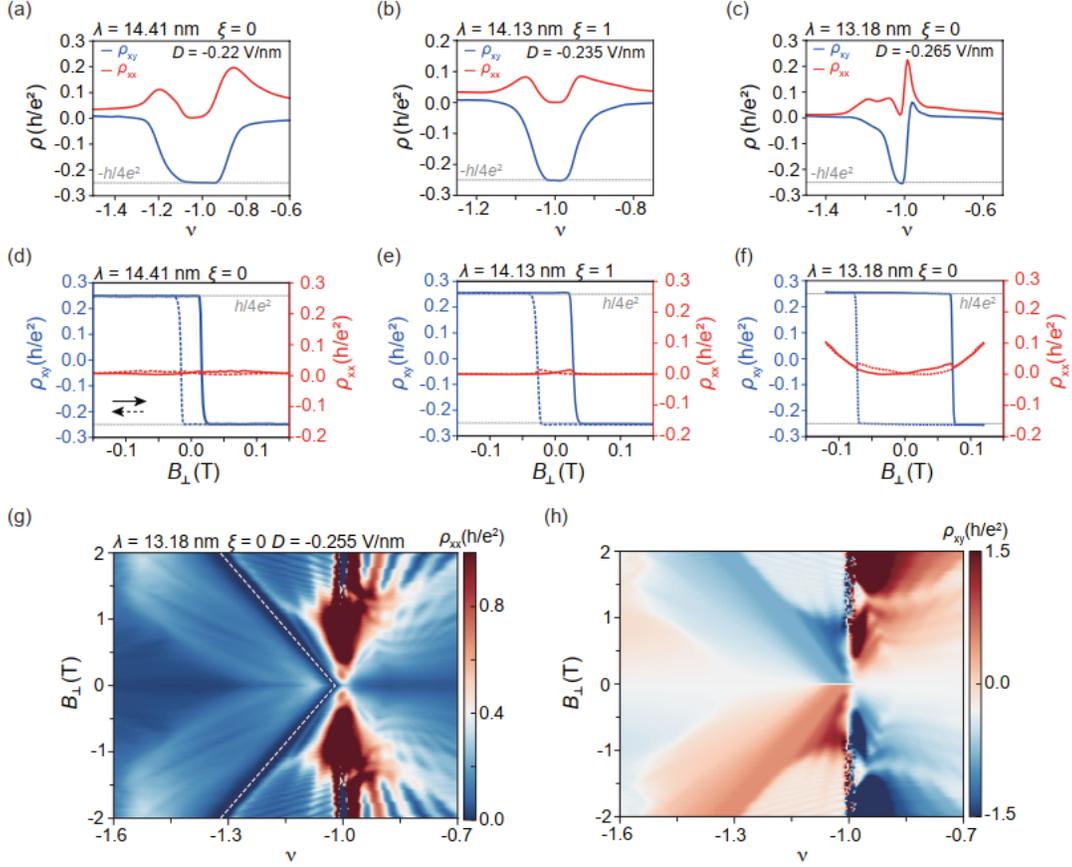

FIG. S3. Transport characterizations of $C = -4$ Chern insulators at $\nu = -1$ in device 1-3. (a)-(c) $\rho_{xx}$ and $\rho_{xy}$ as a function of moiré filling $\nu$ at fixed $D$, measured at $B_\perp = 0.1$ T and $T = 15$ mK. Data are shown for devices with different twist angles and alignment orientations, and the corresponding $D$ values are labeled in the respective panels. Data in panels (a) and (c) are symmetrized ($\rho_{xx}$) and anti-symmetrized ($\rho_{xy}$) at $B_\perp = \pm 0.1$ T, respectively. (d)-(f) Magnetic hysteresis loops at $\nu = -1$ and $T = 15$ mK with fixed $D$ of -0.228 V/nm (d), -0.215 V/nm (e), and -0.255 V/nm (f) for different devices. Device details are indicated in each panel. (g) and (h) Landau fan diagram of $\rho_{xx}$ (g) and $\rho_{xy}$ (h) measured in device 3 at $D = -0.255$ V/nm and $T = 15$ mK. The white dashed lines represent slopes corresponding to Chern numbers $|C|$ according to the Streda formula, and align well with the $C = -4$ Chern state at $\nu = -1$.

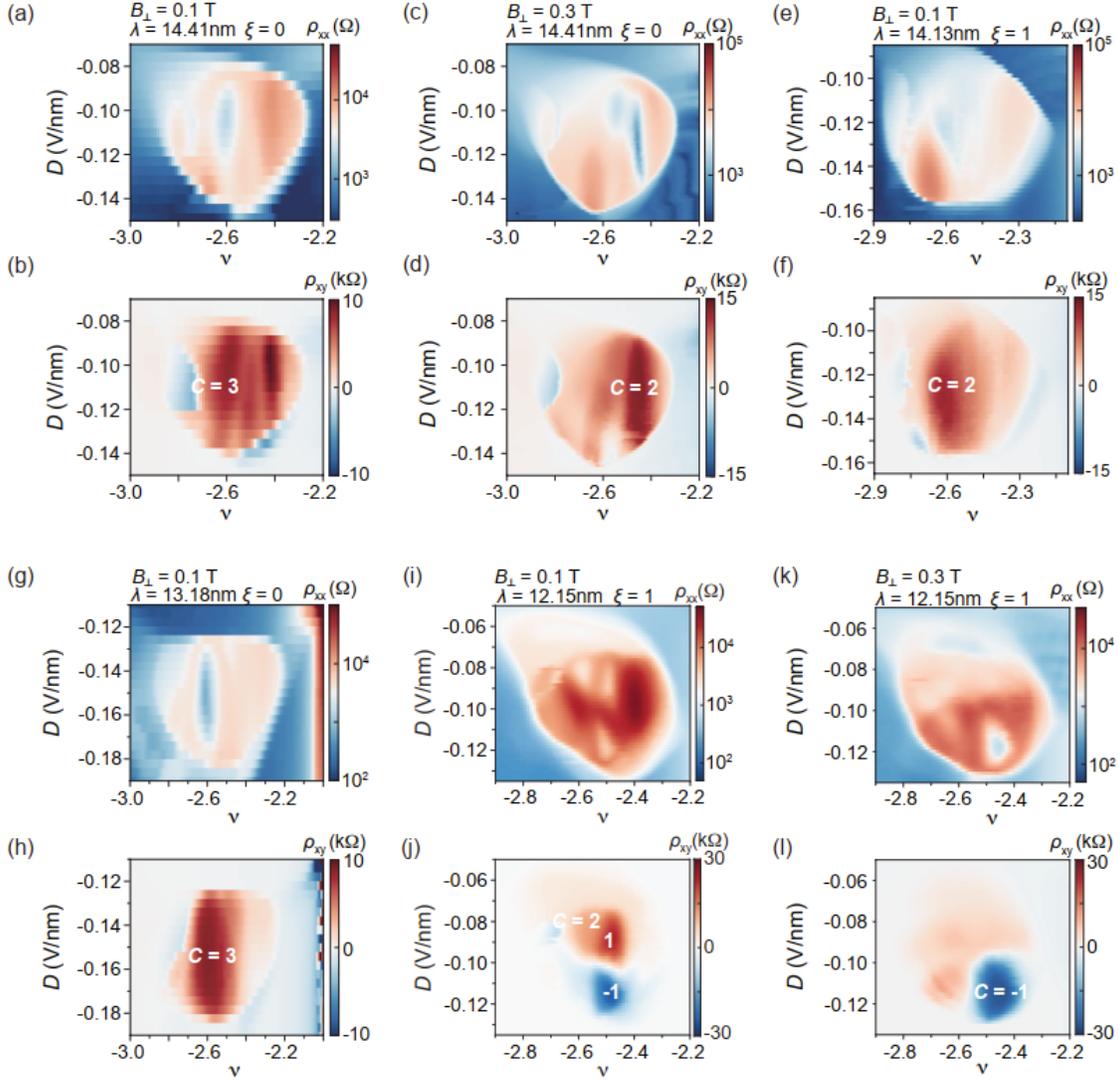

FIG. S4. Tunable high Chern number states near $\nu = -2.5$ in various devices. (a) and (b) Phase diagram of $\rho_{xx}$ (a) and $\rho_{xy}$ (b) as functions of $\nu$ and $D$ for device 1 (with $\lambda = 14.41$ nm), measured at $B_\perp = 0.1$T and $T = 15$ mK. (c) and (d) Analogous measurements to panels (a) and (b) but measured at $B_\perp = 0.3$T. (e) and (f) Phase diagram of symmetrized $\rho_{xx}$ (e) and anti-symmetrized $\rho_{xy}$ (f) as functions of $\nu$ and $D$ for device 2 (with $\lambda = 14.13$ nm), measured at $B_\perp = \pm 0.1$T and $T = 15$ mK. (g) and (h) Phase diagram of symmetrized $\rho_{xx}$ (a) and anti-symmetrized $\rho_{xy}$ (b) as a function of $\nu$ and $D$ for device 3 (with $\lambda = 13.18$ nm), measured at $B_\perp = \pm 0.1$T and $T = 15$ mK. (i) and (j) Phase diagrams of symmetrized $\rho_{xx}$ (i) and anti-symmetrized $\rho_{xy}$ (j) as a function of $\nu$ and $D$ for device 5 ($\lambda = 12.15$ nm), measured at $B_\perp = \pm 0.1$T and $T = 15$ mK. (k) and (l) Analogous measurements to panels (i) and (j) but measured at $B_\perp = 0.3$T. High-Chern-number Chern insulators are observed across all devices, exhibiting different Chern numbers that are highly tunable as a function of twist angles, displacement electric field, and magnetic field.

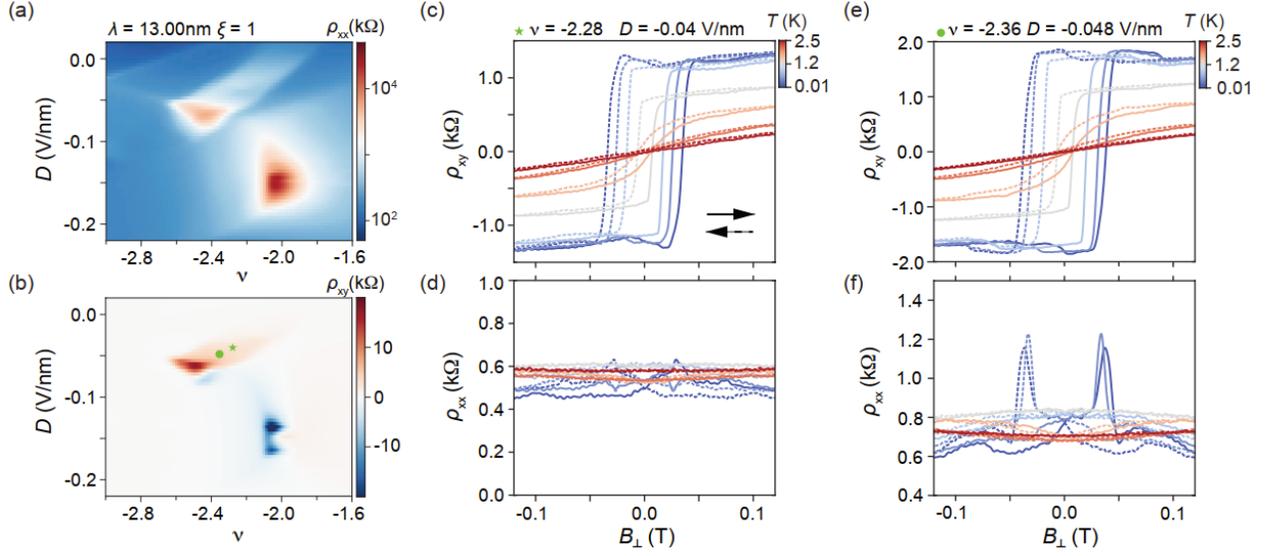

FIG. S5. Anomalous Hall effect near $\nu = -2.5$ in device 4 ($\lambda = 13$ nm). (a) and (b) Phase diagram of symmetrized $\rho_{xx}$ (a) and anti-symmetrized $\rho_{xy}$ (b) as a function of $\nu$ and $D$ measured at $B_\perp = \pm 0.1$ T and $T = 15$ mK. (c)-(f) Temperature-dependent magnetic hysteresis loops of $\rho_{xy}$ and $\rho_{xx}$ measured at $\nu = -2.28$, $D = -0.04$ V/nm (c, d) and $\nu = -2.36$, $D = -0.048$ V/nm (e, f), which are also indicated in panel (b). Obviously, an anomalous hall effect with spontaneous time reversal symmetry breaking is observed near $\nu = -2.5$.

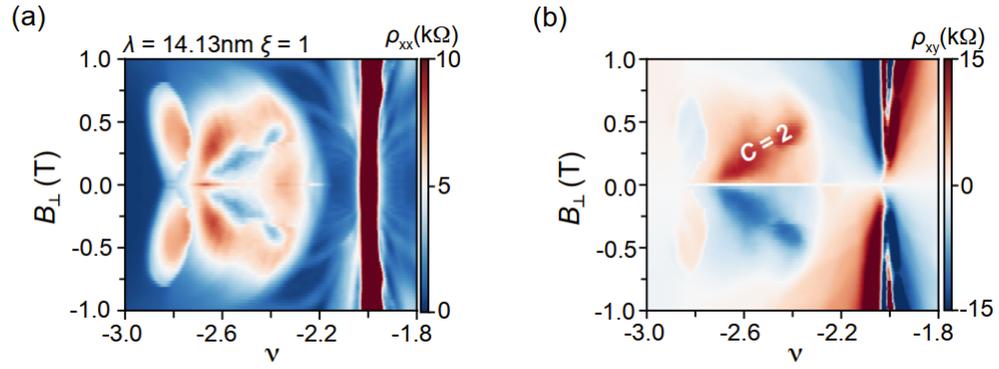

FIG. S6. The emergence of $C = 2$ Chern insulators near $\nu = -2.5$ in device 2 ($\lambda = 14.13$ nm). (a) and (b) Landau fan diagrams of $\rho_{xx}$ (a) and $\rho_{xy}$ (b) measured at $D = -0.125$ V/nm and $T = 15$ mK.

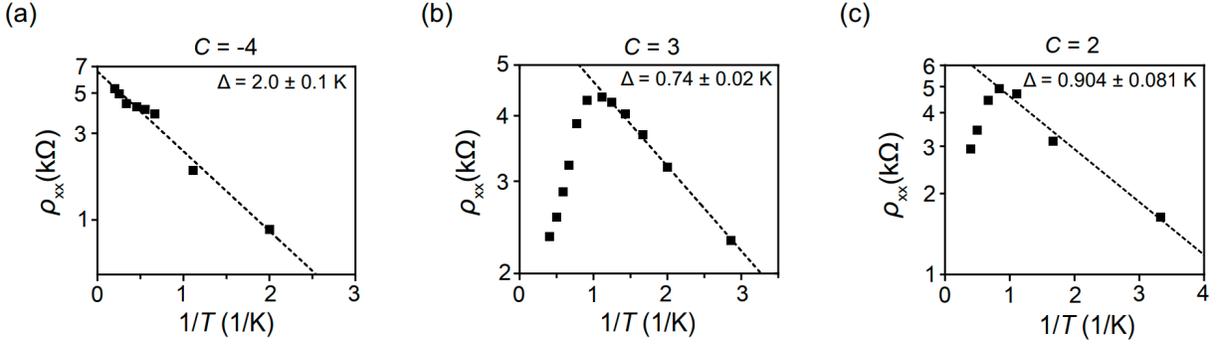

FIG. S7. Thermal activation gaps for different Chern states. (a)-(c) Arrhenius plots of $\rho_{xx}$ for the $C = -4$ state measured at $v = -1$ and $D = -0.228$ V/nm in device 1 (a); the $C = 3$ state measured at $v = -2.6$ and $D = -0.12$ V/nm in device 3 (b); and the $C = 2$ state measured at $v = -2.6$ and $D = -0.081$ V/nm in device 5 (c).

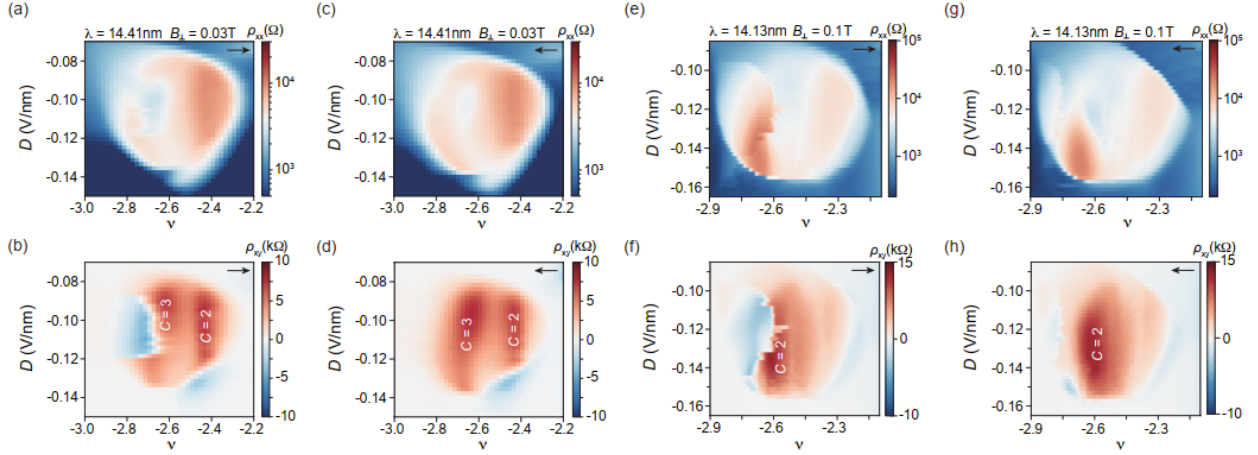

Fig. S8. Additional characterization for the electric switching of magnetic moments for Chern states near $\nu = -2.6$ in device 1 and 2. (a) and (b) Phase diagrams of $\rho_{xx}$ (a) and $\rho_{xy}$ (b) as a function of $\nu$ and $D$ in device 1, measured at $B_\perp = 0.03$ T with the density swept in the upward direction. (c) and (d) Same measurements as in (a) - (b), but with the density swept in the downward direction. (e)-(h) Analogous measurements to those in (a)-(d), measured in device 2 at $B_\perp = \pm\, 0.1$ T. All measurements were performed at $T = 15$ mK. Therefore, together with Fig. S4(g) and (h), where the density is swept in the upward direction, the results indicate that electric-field switching of magnetic moments for Chern states near $\nu = -2.6$ is observed in devices 1–3.

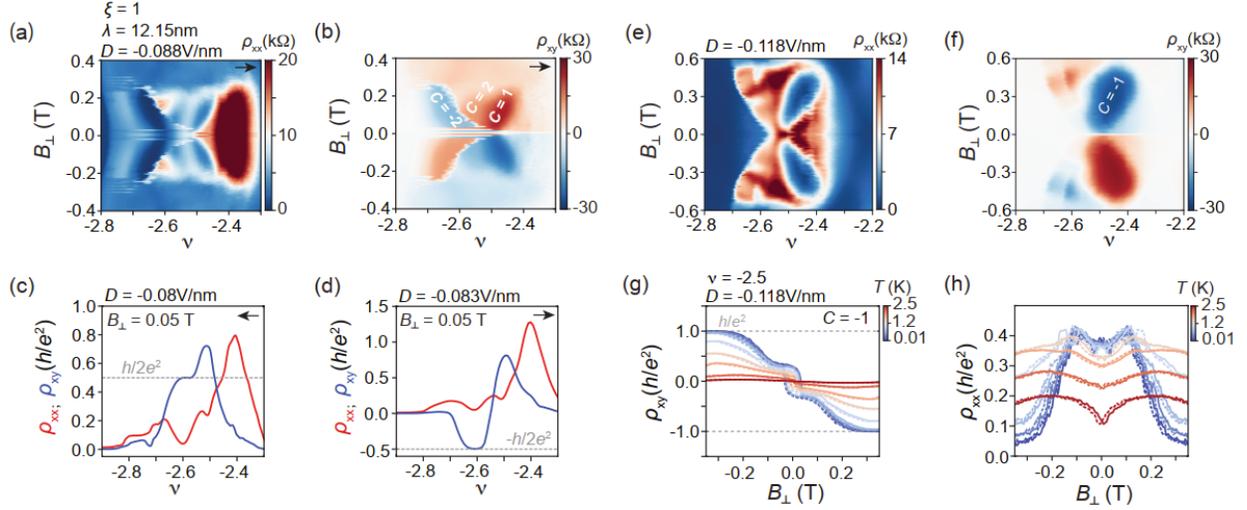

FIG. S9. Additional data for the Chern states emerging near $\nu=-2.5$ and $-2.6$ in device 5 ($\lambda = 12.15$ nm). (a) and (b) Same as Fig. 4(e)-4(f), but with the density swept upward; in Fig. 4(e)–4(f), the density is swept downward. (c) and (d) Symmetrized $\rho_{xx}$ and antisymmetrized $\rho_{xy}$ as a function of $\nu$ measured at $B_\perp = \pm 0.05$T, with $D = -0.08$ V/nm (c) and $D = -0.083$ V/nm (d), respectively. (e) and (f) Landau fan diagram of $\rho_{xx}$ (e) and $\rho_{xy}$ (f) measured at $D = -0.118$ V/nm. It is noteworthy that the electric switching behovior is absent for the $C = \pm 1$ state at $\nu = -2.5$, and the Landau fan diagram is the same for both sweeping directions. (g) and (h) Temperature dependence of the magnetic hysteresis loops for $\rho_{xy}$ (g) and $\rho_{xx}$ (h), measured at $\nu = -2.5$ and $D = -0.118$ V/nm, for the $C = -1$ Chern state. All panels except (g) and (h) were measured at $T = 15$ mK.

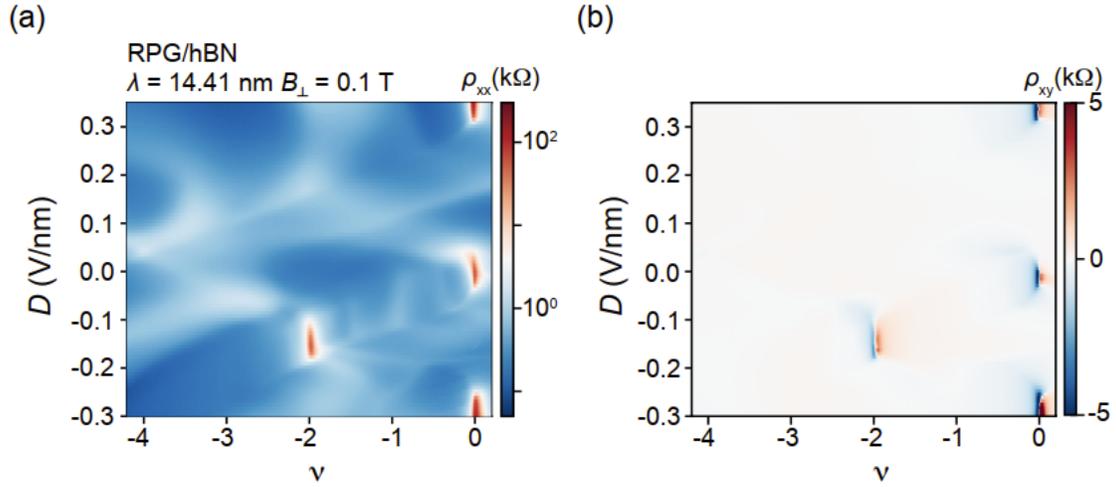

FIG. S10. Transport characterizations of RPG/hBN moiré superlattices. (a) and (b) Phase diagram of $\rho_{xx}$ (a) and $\rho_{xy}$ (b) as a function of $D$ and $\nu$ for an RPG/hBN device with a twist angle of 0.2° ($\lambda$ = 14.41 nm; Device 6), measured at $B_\perp$ = 0.1 T and $T$ = 1.7 K. In contrast to RTG/hBN devices with comparable moiré wavelengths (Device 1 and 2), we find no evidence of an anomalous Hall effect or Chern insulators at $\nu$ = -1 or near $\nu$ = -2.5 in this RPG/hBN device.